\documentstyle[12pt]{article}
\begin{document}
\setlength{\oddsidemargin}{0cm}
\setlength{\baselineskip}{7mm}

\def\del{\partial}
\def\pbar{\bar p}
\def\xbar{\bar x}
\def\unit{{\bf 1}}
\def\picunit#1{{\bf 1}^{#1}}
\def\dualunit#1{{\bf 1}_{*}^{#1}}
\def\singular{\mbox{\bf S}}
\def\ie{{\em i.e., }}
\def\beq{\begin{equation}}
\def\eeq{\end{equation}}
\def\beqa{\begin{eqnarray}}
\def\eeqa{\end{eqnarray}}
\def\vac#1{|#1\rangle}
\def\upvac#1{|#1, \uparrow \rangle}
\def\fock#1{{\cal F}_#1}

\begin{titlepage}

    \begin{normalsize}
     \begin{flushright}
                 UT-Komaba/93-23 \\
                 November 1993
     \end{flushright}
    \end{normalsize}
    \begin{LARGE}
       \vspace{1cm}
       \begin{center}
         Note on $N=0$ string as $N=1$ string \\
       \end{center}
    \end{LARGE}

  \vspace{5mm}

\begin{center}
           Hiroshi I{\sc shikawa}
           \footnote{E-mail address:
              ishikawa@hep1.c.u-tokyo.ac.jp}
                \ \  and \ \
           Mitsuhiro K{\sc ato}
           \footnote{E-mail address:
              katom@tkyvax.phys.s.u-tokyo.ac.jp} \\
      \vspace{4mm}
        {\it Institute of Physics, College of Arts and Sciences} \\
        {\it University of Tokyo, Komaba}\\
        {\it Meguro-ku, Tokyo 153, Japan}\\
      \vspace{1cm}

    \begin{large} ABSTRACT \end{large}
        \par
\end{center}
\begin{quote}
 \begin{normalsize}
\ \ \ \
A similarity transformation, which brings a particular class of the
$N=1$ string to the $N=0$ one, is explicitly constructed. It enables
us to give a simple proof for the argument recently proposed by
Berkovits and Vafa.
The $N=1$ BRST operator is turned into the direct sum of
the corresponding $N=0$ BRST operator and that for an additional topological
sector. As a result, the physical
spectrum of these $N=1$ vacua is shown to be isomorphic to
the tensor product of the $N=0$ spectrum and the topological sector
which consists of only the vacuum.
This transformation manifestly keeps the operator algebra.

 \end{normalsize}
\end{quote}

\end{titlepage}
\vfil\eject


The aim of this short note is to give a simple proof to the argument
initiated by Berkovits and Vafa\cite{BV} that the $N=0$ (bosonic)
string can be regarded as a special case of the $N=1$ (fermionic)
string.

The idea of ref.~\cite{BV} is the following. For any given $c=26$ matter
with its energy-momentum tensor $T_m(z)$, we have $c=15$ ($\hat c=10$) system
by coupling spin $(3/2,-1/2)$ fermionic ghosts which we denote as $(b_1, c_1)$.
Then this system can be considered as a matter system of critical
$N=1$ fermionic string because of the existence of $N=1$ super
conformal algebra in the system:
\beqa\label{N=1op}
  T_{N=1} &=& T_m - \frac{3}{2} b_1 \del c_1 - \frac{1}{2} \del b_1 \, c_1
        + \frac{1}{2} \del^2 (c_1 \del c_1) \,,\\
  G_{N=1} &=& b_1 + c_1 (T_m + \del c_1 \, b_1) + \frac{5}{2} \del^2 c_1
  \,,
\eeqa
which satisfy
\beqa
  T_{N=1}(z) T_{N=1}(w) & \sim & \frac{15/2}{(z-w)^4}
        + \frac{2}{(z-w)^2} T_{N=1}(w) + \frac{1}{z-w} \del T_{N=1}(w) \,,\\
  T_{N=1}(z) G_{N=1}(w) & \sim & \frac{3/2}{(z-w)^2} G_{N=1}(w)
                        + \frac{1}{z-w} \del G_{N=1}(w) \,,\\
  G_{N=1}(z) G_{N=1}(w) & \sim & \frac{10}{(z-w)^3} +
       \frac{2}{z-w} T_{N=1}(w)  \,.
\eeqa
Tensoring spin $(2,-1)$ reparametrization ghosts $(b,c)$ and spin
$(3/2,-1/2)$ super ghosts $(\beta , \gamma)$ to the above, we can
construct whole Fock space of $N=1$ fermionic string
\begin{equation}\label{N=1space}
  \fock{{N=1}} = (T_m) \bigotimes (b_1, c_1) \bigotimes (b, c) \bigotimes
  (\beta, \gamma)
\,.
\end{equation}
Now their claim is that this $N=1$ string\footnote[1]{Hereafter we
  mean this special class of $N=1$ string by simply saying $N=1$
  string.} is equivalent to the $N=0$ bosonic string with the matter
$T_m(z)$ after integrating out the $(b_1,c_1)$ and $(\beta , \gamma)$.
The equivalence stands not only for the reduction of the Fock space
(\ref{N=1space}) to the $N=0$ Fock space
\begin{equation}
  \fock{{N=0}} = (T_m) \bigotimes (b, c)
\,,
\end{equation}
but also for the one to one correspondence between the physical
observables of each theory such as scattering amplitude.

The authors of ref.~\cite{BV} gave a plausible argument for how $N=1$
amplitude is reduced to the corresponding one in the $N=0$ for the
case of the physical vertex made from the spin 1 operator of the
matter in $T_m(z)$. But generic physical vertex operators are
expressed as mixed form of matter and ghosts, {\it e.g.} ground ring
operators in the two dimensional string. So more general proof is
desired to establish the equivalence.

Ref.~\cite{F} is an attempt toward this end, but is still insufficient
as the author himself admitted. In particular, it is not clear whether
the operator product algebras of the physical vertex operators in
$N=1$ and $N=0$ are isomorphic each other, which is must for the
coincidence of amplitude.

In the following we will construct a simple similarity transformation
which maps the $N=1$ theory to the tensor product of the $N=0$ theory and a
trivial topological
sector whose physical state is only its vacuum. The $N=1$ BRST charge
is mapped to the direct sum of the one in the $N=0$ theory and the one
in the topological sector, each of which acts on its own sector
exclusively. Therefore the relation between the BRST cohomology of the
$N=1$ and that of the $N=0$ is very clear as will be shown.
Moreover, algebra isomorphism is manifest due to the similarity
transformation. This is exactly the same technique used in
ref.~\cite{IK}
in which present authors have shown that the equivalence of BRST
cohomologies of $SL(2,R)/U(1)$ black hole and $c=1$ string.

Let us start with the definition of the BRST charge for the $N=1$
string obtained from $c=26$ matter in the above:
\begin{eqnarray}
  Q_{N=1} &=& \oint\! dz \left(
        c T_{N=1} - \frac{1}{2} \gamma G_{N=1} + b c \del c
                - \frac{1}{4} b \gamma^2
        + \frac{1}{2} \del c \, \beta \gamma - c \beta \del \gamma
                \right) \nonumber \\
          &=&   \oint\! dz
        \left[
         \left(  c - \frac{1}{2} \gamma c_1 \right) T_m
        + c \left( - \frac{3}{2} b_1 \del c_1 - \frac{1}{2} \del b_1 \, c_1
        + \frac{1}{2} \del^2 (c_1 \del c_1) \right) \right. \nonumber \\
           & &\left.
        -\frac{1}{2} \gamma \left(
                b_1 + b_1 c_1 \del c_1 + \frac{5}{2} \del^2 c_1
        \right)
        + b c \del c - \frac{1}{4} b \gamma^2
        + \frac{1}{2} \del c \, \beta \gamma - c \beta \del \gamma\,
        \right]
\,.
\end{eqnarray}
The physical states of the $N=1$ string are defined as $Q_{N=1}$ cohomology.
Due to the coupled terms of the field in $\fock{{N=0}}$ and the fields
in the topological sector
$\fock{{top}} = (b_1, c_1) \bigotimes (\beta, \gamma)$,
it is not so trivial to show that the $Q_{N=1}$
cohomology is reduced to the $N=0$ one by means of this form of
$Q_{N=1}$.

Our main result is the following similarity transformation which maps
$Q_{N=1}$ into very simple form:
\begin{equation}\label{transf}
  e^R Q_{N=1} e^{-R} = Q_{N=0} + Q_{top} \,,
\end{equation}
where
\begin{equation}
    R = \oint\! dz\, c_1 \left(
        \frac{1}{2} \gamma b - 3 \del c \, \beta - 2 c \del \beta
        - \frac{1}{2} \del c_1 c b
        + \frac{1}{4} \beta \gamma \del c_1 \right) \,,
\end{equation}
and
\begin{equation}
  Q_{N=0} = \oint\! dz \left( c T_m + b c \del c \right) \,,
\end{equation}
\begin{equation}
  Q_{top} = \oint\! dz \left( -\frac{1}{2} b_1 \gamma \right) \,.
\end{equation}
$Q_{N=0}$ is the BRST operator in the $N=0$ string which acts
only on $\fock{{N=0}}$ and $Q_{top}$ is the one in the topological sector
which also acts only on its own sector $\fock{{top}}$.
Now, with this form of the BRST operator, it is clear that $Q_{N=1}$
cohomology is direct product of $Q_{N=0}$ cohomology and $Q_{top}$
cohomology.
And $Q_{top}$ cohomology only consists of the vacuum in $\fock{{top}}$,
because $b_1$, $\gamma$ are $Q_{top}$-exact and $c_1$, $\beta$ are not
$Q_{top}$-closed so that all the states in $\fock{{top}}$ but the
vacuum are unphysical or $Q_{top}$-exact. Thus we obtain one to one
correspondence between the $Q_{N=1}$ and $Q_{N=0}$ cohomologies. Since
(\ref{transf}) is similarity transformation, it manifestly keeps the
operator algebra. As a result, any correlation functions in the $N=1$
theory are reduced to the corresponding ones in the $N=0$ theory.
These establish the equivalence of $N=0$ string and a special class of
$N=1$ string constructed from $\fock{{N=1}}$.

It may be helpful to see how $N=1$ physical vertex operators are
mapped to corresponding ones in $N=0$ theory in our transformation.
For example, a vertex operator in the ghost number one picture
treated in \cite{BV}, which is expressed as
$c V - {1 \over 2}\gamma c_1 V$ in our notation, is mapped as
\begin{equation}
  e^R (c V - {1 \over 2}\gamma c_1 V) e^{-R} = c V \,.
\end{equation}
The right hand side is desired form in the $N=0$ string. On the other
hand, the ghost number zero picture operator is invariant under the
transformation, so that n-point amplitude on a sphere satisfies
\begin{eqnarray}
&&\langle(c\ e^{-\phi}c_1V_1)  (c\ e^{-\phi}c_1V_2)
  (c\ V_3 - {1 \over 2}\gamma c_1 V_3)
  \int V_4 ...\int V_n \rangle_{\fock{{N=1}}} \nonumber \\
&&= \langle (c\ V_1)  (c\ V_2) (c\ V_3) \int V_4 ...\int V_n
  \rangle_{\fock{{N=0}}} \times \langle (e^{-\phi}c_1)(e^{-\phi}c_1)
  \rangle_{\fock{{top}}} \,.
\end{eqnarray}
The second factor in the right hand side is just a trivial factor
which is usually taken to be one. These argument, of course, can be
apparently extended to any vertex operators in any pictures using our
transformation.


Some remarks are in order.

(a) As is seen from eq.(\ref{N=1op}), the energy-momentum tensor of the
$N=1$ theory has a non-standard form for the $(b_1, c_1)$-sector; modified by
a total derivative term made of $c_1$. However, this term
disappears after the transformation and the energy-momentum tensor of
$N=0$ and topological sector turn into the standard ones
\beq
  e^R T_{total} e^{-R} = T_m - 2 b \del c - \del b \, c
        - \frac{3}{2} b_1 \del c_1 - \frac{1}{2} \del b_1 \, c_1
        - \frac{3}{2} \beta \del \gamma - \frac{1}{2} \del \beta \, \gamma\, ,
\eeq
where we denote the total energy-momentum tensor of the $N=1$ string as
$T_{total}$.

(b) After the transformation by $R$, the physical states take the
form $|phys \rangle_{N=0} \bigotimes |0 \rangle_{top} $.
Since $R$ has only ($b_1,c_1,\beta,\gamma$)-ghost number even
terms, the transformation does not change even-odd property with
respect to this ghost number (let us denote it by $N_F$).
Therefore, it is obvious from $R$-transformed expression that all the
physical states are $N_F$ even.

(c) GSO projection can be defined with this $N_F$ in the $N=1$
theory.
The physical spectrum consists of $N_F$ even states
only, as is seen from the above. So, the physical contents of the theory would
be empty if we choose the $N_F$ odd sector in the projection.
In the Ramond sector, similarly, the sector with the same $N_F$ parity
as the ground state
has to be singled out.

\vspace{0.5in}
\section*{Acknowledgements}
We have been much benefitted from {\tt ope.math} the package for the
operator product expansion developed by A.~Fujitsu.
The research of M.K. is supported in part by
the Grant-in-Aid for Encouragement of Young Scientists (\# 05740168)
and by that for Scientific Research on Priority Areas ``Infinite Analysis''
(\# 05230011) from the Ministry of Education, Science and Culture.



\begin{thebibliography}{99}

\bibitem{BV} N.~Berkovits and C.~Vafa,
{\em On the Uniqueness of String Theory},
preprint HUTP-93/A031, KCL-TH-93-13, hep-th/9310170 (October 1993).

\bibitem{F} J.~M.~Figueroa-O'Farrill,
{\em On the Universal String Theory},
preprint QMW-PH-93-29, hep-th/9310200 (October 1993).

\bibitem{IK} H.~Ishikawa and M.~Kato, Phys.~Lett. {\bf B302} (1993) 209.

\end{thebibliography}
\end{document}